\documentclass[twocolumn,showpacs,preprintnumbers,amsmath,amssymb]{revtex4}

\usepackage{graphicx}
\usepackage{dcolumn}
\usepackage{bm}
\usepackage{psfrag}

\def\vec#1{{\bf #1}}
\def\ex{\vec{e}_x}
\def\ey{\vec{e}_y}
\def\ez{\vec{e}_z}
\def\ui{^{(i)}}

\def\he{\vec{h}_{\rm eff}}
\def\mza{m^{(1)}_z}
\def\mzb{m^{(2)}_z}

\def\vmo{\vec{m}_0}
\def\vmui{\vec{m}_1^{(i)}}

\begin{document}

\preprint{APS/123-QED}

\title{Dynamic Domains\\
in Strongly Driven Ferromagnetic Films}

\author{K. Mayes}
\author{T. Plefka}%
 \email{timm@arnold.fkp.physik.tu-darmstadt.de}
\author{H. Sauermann}%
\affiliation{%
Theoretische Festk\"orperphysik, Technische Universit\"at Darmstadt, D-64289 Darmstadt, Germany
}%


\date{\today}

\begin{abstract}
The spatiotemporal structure formation problem is investigated in
the region far above the transverse ferromagnetic resonance
instability. The investigations are based on the dissipative
Landau-Lifshitz equation and have been performed on a model which
takes external fields, isotropic exchange fields, anisotropy
fields and the demagnetizing part of the dipolar field into
consideration. The numerical simulations for these models exhibit
stationary domain structure in the rotating frame. Employing
analytical methods and simplifying  the model, certain features,
such as the magnetization within the domains and the proportion of
the system in each domain, are described analytically.
\end{abstract}
\pacs{76.50+g, 75.60.Ch}
\maketitle

\section{Introduction}

Ferromagnetic systems driven by external magnetic fields have been
under investigation for many years \cite{Damon}. Early work by
Suhl already showed that there is a threshold of the pump
amplitude  where the uniform state becomes unstable to homogeneous
driving fields \cite{Suhl1957}. When the difference between the
pump field and its critical value is small (the weakly nonlinear
case), the experimental findings can be analyzed by extensions of
the theory of  Suhl as reviewed in \cite{wigen}. The pattern
formation in such a system can be described by amplitude equations
\cite{el,FM1996}.

However, when the probe is {\it strongly} driven, such
perturbative procedures break down. One-dimensional numerical
simulations have predicted dynamic domains for a model including a
transversely rotating field \cite{Elme1988}. Dynamic domains are
stable solutions to the equations of motion. In the reference
frame rotating with the driving field they exhibit a stationary
domain structure, like that known from static domains, yet in the
lab frame their position is stationary while the magnetization
within the domains rotates at different angles. Plefka also
obtained dynamic domains in one dimension by numerical simulation
and was able to explain characteristic elements of the structure
analytically \cite{Plef1995} for a very simplified model.

The experimental difficulties involved with large pump amplitudes
were solved already twenty years ago \cite{Doetsch}. Renewed
interest in this topic has led to recent work, where, using the
Faraday effect, dynamic domains were observed in garnet films
driven by high power inhomogeneous driving fields \cite{Woe1998}.

The length scales of the patterns we are interested in are large
compared to the atomic distance of such substances and so the
relevant quantity is a macroscopic variable, the local
magnetization $\vec{m}(\vec{r},t)$. Using a model that includes a
saturating static field and a strong transverse pump field
\cite{Mayes}, we consider the structures occurring in a
ferromagnetic film of side $L$ with normal in the $z$-direction.

 The paper is organized as following: In Sec.II the physical model
 is described which enters in the Landau-Lifshitz Equation via the
 effective fields. The numerical results for the various dynamical domains
 are presented in Sec. III . Employing  simplifications of the original model
the linear stability of the structures found is considered in
 Sec. IV leading to partial description of the numerical
 results. Conclusions are presented in Sec.V.

\section{Landau-Lifshitz Equation}

The dynamics of $\vec{m}(\vec{r},t)$ are described by the
Landau-Lifshitz equation \cite{LaLi1935,Plef93}:
\begin{equation}
\partial_t\,{\vec{m}}=-\vec{m}\times\vec{h}_{\rm eff}-
\Gamma\vec{m}\times(\vec{m}\times\vec{h}_{\rm eff})\quad .\label{ll1}
\end{equation}
$\Gamma$ is a dimensionless damping coefficient.
The effective magnetic field  $\vec{h}_{\rm eff}$ is made up of both external and
internal magnetic fields:
\begin{eqnarray}\label{ll2}
\vec{h}_{\rm eff}&=& H\vec{e}_z + h\left( \cos(\omega t)\vec{e}_x
+\sin(\omega t) \ey \right)\\\nonumber
 && + J \nabla^2\vec{m} - \overline{m}_z
\vec{e}_z + K m_z \vec{e}_z \quad .
\end{eqnarray}
These terms are, in order of appearance: a static field of magnitude $H$
in the $z$-direction (perpendicular to the plane of the
film); an in-plane pump field with amplitude $h$ and frequency $\omega$;
an isotropic exchange
field ($J>0$); the demagnetizing part of the dipolar field, describing the
geometry of the film; and a uniaxial
anisotropy, also in the $z$-direction.

The explicit time dependence of (\ref{ll1}) is removed by a transformation to
the rotating reference frame, yielding
\begin{equation}\label{eq1}
\partial_t\,{\vec{m}}=-\vec{m}\times(\vec{h}_{\rm eff}-\omega \ez)-
\Gamma\vec{m}\times(\vec{m}\times\vec{h}_{\rm eff})\quad ,
\end{equation}
while the effective magnetic field now has
the form
\begin{equation}\label{eq2}
\vec{h}_{\rm eff}= H \vec{e}_z+h \vec{e}_x + J\nabla^2\vec{m}-\overline{m}_z
\vec{e}_z+K m_z \vec{e}_z\quad .
\end{equation}

An alternative, frequently used form of the Landau-Lifshitz equation is the Gilbert form
\begin{eqnarray}\label{ll5}
&&\partial_t \,{\vec{m}}-\Gamma \vec{m} \times \partial_t \, \vec{m}=\\
&&-(1+\Gamma^2)\,
\vec{m}\times \left[\vec{h}_{\rm
eff}-\frac{\omega}{1+\Gamma^2}\vec{e}_z +\frac{\Gamma
\omega}{1+\Gamma^2}\vec{m} \times \ez \right] \,\, .\nonumber
\end{eqnarray}
This is obtained by multiplying equation (\ref{eq1}) from the left by $\Gamma \vec{m}
\times$ and recalling that $\vec{m}\cdot \vec{m}=1$.
The damping coefficient $\Gamma$ is the same in both the Gilbert
form of the Landau-Lifshitz equation (\ref{ll5}) and the common
form (\ref{eq1}).

Throughout this work we present results in the rotating reference frame. We recall that
static results in this frame of reference transform to structures rotating with
frequency $\omega$ in the $x$-$y$ plane in the laboratory frame. Hence
a state that is {\it statically} stable in the rotating
reference frame is {\it dynamically} stable in the lab frame.

\section{Numerical Simulation}

\begin{figure}
\includegraphics[width=0.4\textwidth]{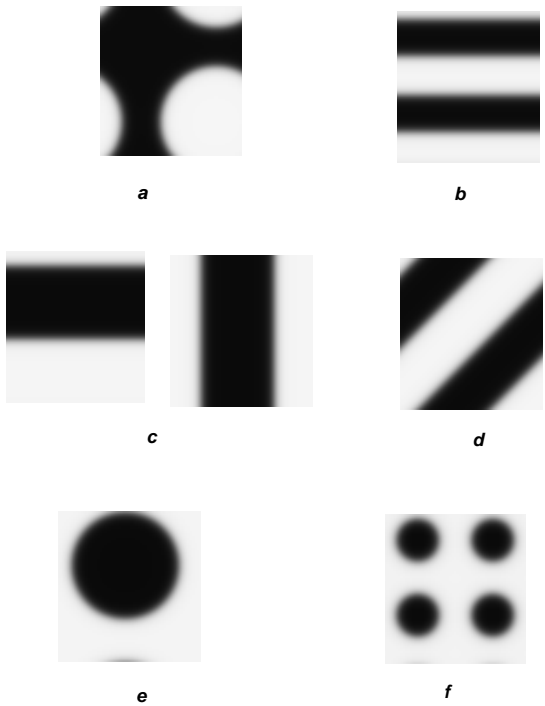}
\caption[]{Dynamically stable domain states obtained at the
same parameter values: $\omega=H=3$, $h=2$, $J=10$, $K=5$, $\Gamma=0.1$. a)
``spin-up'' bubble, period $L$; b) horizontal stripes, period
$L/2$; c) horizontal and vertical stripes, period $L$; d) diagonal
stripes, period $\sqrt{L/2}$; e) ``spin-down'' bubble, period
$L$; f) ``spin-down'' bubble, period $L/2$} \label{ndd1}
\end{figure}

We simulate equation (\ref{eq1}) with (\ref{eq2}) on a
vector-parallel computer and workstations, using a spectral
approach to deal with the internal field contributions. The number
of mesh points for the two-dimensional system was varied up to
$256 \times 256$ and it is seen that a minimum spatial grid size
of $64\times 64$ is necessary to recognize to good accuracy all
magnetic structures that are also visible with a larger number of
mesh points. Periodic boundary conditions are used. For $64\times
64$ mesh points, the boundary conditions also have no effect on
the structures within the system. This is confirmed by also
performing the simulations with
unpinned spins, and observing that the same results are obtained.
A Euler integration scheme is sufficient and therefore used for
the time integration of the system. Indeed more advanced methods,
like  Runge-Kutta and NDSolve of Mathematica, give essentially
identical numerical results.

Throughout the course of a simulation we require that the
magnitude of  the magnetization remain constant $|\vec{m}|=1$.
However, the accumulation of numerical errors means that this is
not the case, and the magnetization tends to drift away from the
value one. Therefore we add an additional Bloch-like damping term
to the equation of motion (\ref{eq1}) that recalls the length of
the magnetization vector back to one at each time step. This Bloch
damping, although
 a reaction to a numerical artifact, is also justified physically \cite{Plef93}.

To investigate systematically the presence of dynamically stable
structures, all numerical and physical parameters except the
amplitude of the driving field are kept constant ($H=\omega=3$,
$J=10$, $K=5$, $\Gamma=0.1$), while $h$ is varied from $h=0$ upwards. Depending on the
initial states and on the value of $h$, different coexisting dynamically stable solutions
are found.

\subsection{Homogeneous State}

The solution found most frequently in numerical simulations is the homogeneous or uniform
solution. In this case, all the spins throughout the system rotate
at the same frequency, in phase, and with the same constant $m_z$.
The dynamically stable homogeneous state is
found numerically for values of $h$
below a certain critical value $h_{c1}\approx 3.7$ and above another critical
value $h_{c2}\approx 4.9$.

\subsection{Stripes and Bubbles}

Some of the two-domain dynamically stable
structures also found are shown in Figure \ref{ndd1}. This is a shaded contour
plot of the $m_z$ component of the magnetization throughout the system. The dark
areas depicted are those where $m_z<0$ while the light areas imply $m_z>0$.
Figure \ref{ndd1} makes no statement about the $x$ and $y$
components of the magnetization, which (apart from at the wall)
are spatially homogeneous and vary periodically in time with period
$\omega$. This implies that the spins within the domains rotate  in the $x$-$y$
plane in phase and at the same frequency.

\begin{figure}
\psfrag{b}{\small \rm bubble}
\psfrag{s}{\small \rm stripe}
\psfrag{o}{\small \rm homogeneous}
\hspace{-2cm}
\includegraphics[width=0.35\textwidth]{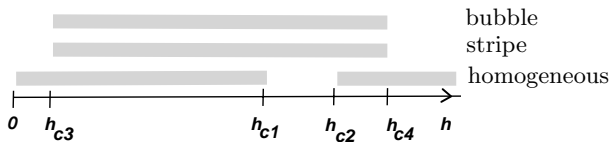}
\caption[]{Schematic view of where the homogeneous,  stripe and bubble
solutions are found numerically, $h$ is the amplitude of the driving field}
\label{ndd3a} \end{figure}

The simulations were performed with a great many different initial conditions.
The results shown in Figure \ref{ndd1} are characteristic of {\it all}
the dynamic domain results obtained, i.e. dynamically stable domain states
either have either translational symmetry ({\bf b, c} and {\bf d}) or
cylindrical symmetry ({\bf a, e} and {\bf f}) in the $z$-component of the magnetization.

\begin{figure*}
\psfrag{m}{\small $m_{z,{\rm max}}-m_{z,{\rm min}}$}
\psfrag{h}{\small $h$}
\psfrag{a}{$h_{c3}$}
\psfrag{b}{$h_{c1}$}
\psfrag{c}{$h_{c2} \approx h_{c4}$}
\psfrag{d}{$h_{c2}$}
\psfrag{e}{$h_{c4}$}
\includegraphics[width=0.7\textwidth]{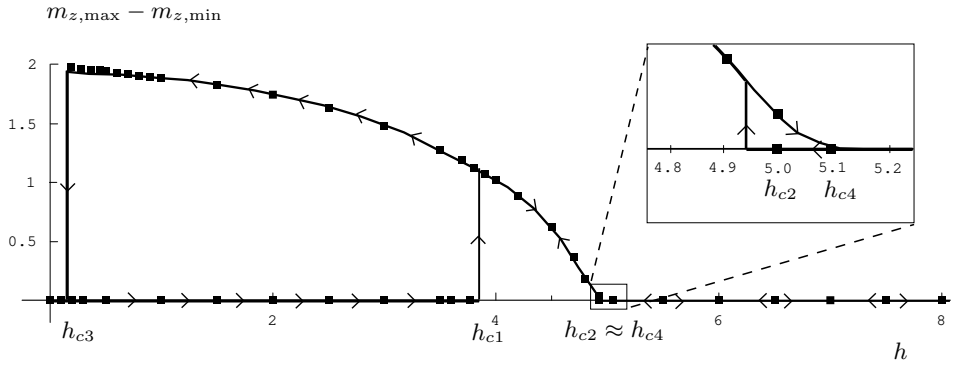}
\caption[]{Hysteresis between the homogeneous and domain
solutions. The figure plots $m_{z,{\rm max}}-m_{z,{\rm min}}$ vs
$h$, the amplitude of the driving field. For the homogeneous state
$m_{z,{\rm max}}-m_{z,{\rm min}} = 0$, while for bubble and stripe
domains $m_{z,{\rm max}}-m_{z,{\rm min}} > 0$. Each dot is the
result of a numerical simulation. Arrows indicate the direction of
change of the driving field $h$ and the direction of transition.
There is a sharp transition from the uniform state to domains at
$h_{c1} \approx 3.7$ and from the domain state to the homogeneous
state at $h_{c3}\approx 0.2$. The region around $h\approx 5$ is
enlarged to show more clearly the sharp transition from the
homogeneous state to the domain state at $h_{c2}\approx 4.9$ and
the continuous transition from the domain state to  the
homogeneous state at the somewhat higher value of $h_{c4}$}
\label{ndd3} \end{figure*}

It is observed that dynamically stable domain states only
exist between two values of the driving field, called $h_{c3}\approx 0.2$ and
$h_{c4}\approx 5$.  As we approach the upper bound $h_{c4}$ the domain
states become ever more flattened out, until eventually at
$h_{c4}$ the domain state merges into the homogeneous state. This
continuous transition is in direct contrast to the lower bound
$h_{c3}$ where a well-defined domain state suddenly no longer
exists. The values of $h_{c3}$ and
$h_{c4}$ are seen to be the same whether bubbles or stripes are
under investigation.

\subsection{Wall Structure}

The spatially distinct uniform regions are separated by a domain
wall. Here the magnetization does not jump from the  value in one
domain to the next, but rather varies continuously across the finite wall width.
The spins inside the wall rotate at the same frequency with the rest of the
system, but with a phase shift compared to the spins within the domains.
In the domain wall all the spins are parallel. In
particular, this is also true for the bubble domain wall and so the whole
structure including wall is {\it not} rotationally symmetric. The rotational
symmetry of the system is broken by the presence of the driving field $h$. Such
bubble domains are always obtained, even if
rotationally symmetric initial conditions are used.

The structure of the wall differs depending on $h$. For small $h$
there is a large negative $m_y$ component of the magnetization.  As $h$
increases, $m_x$ dominates ever more at the center of the wall. In addition,
 the width of the wall increases with $h$. We noted above that the
domains themselves become flattened out. This, together with the
ever widening wall, means that it is difficult to say precisely
when the domain state merges into the homogeneous state at the
upper bound $h_{c4}$.

\subsection{Hysteresis}

\begin{figure}
\psfrag{a}{$t=0$}
\psfrag{b}{$t=20 T$}
\psfrag{c}{$t=30 T$}
\psfrag{d}{$t=40 T$}
\includegraphics[width=0.4\textwidth]{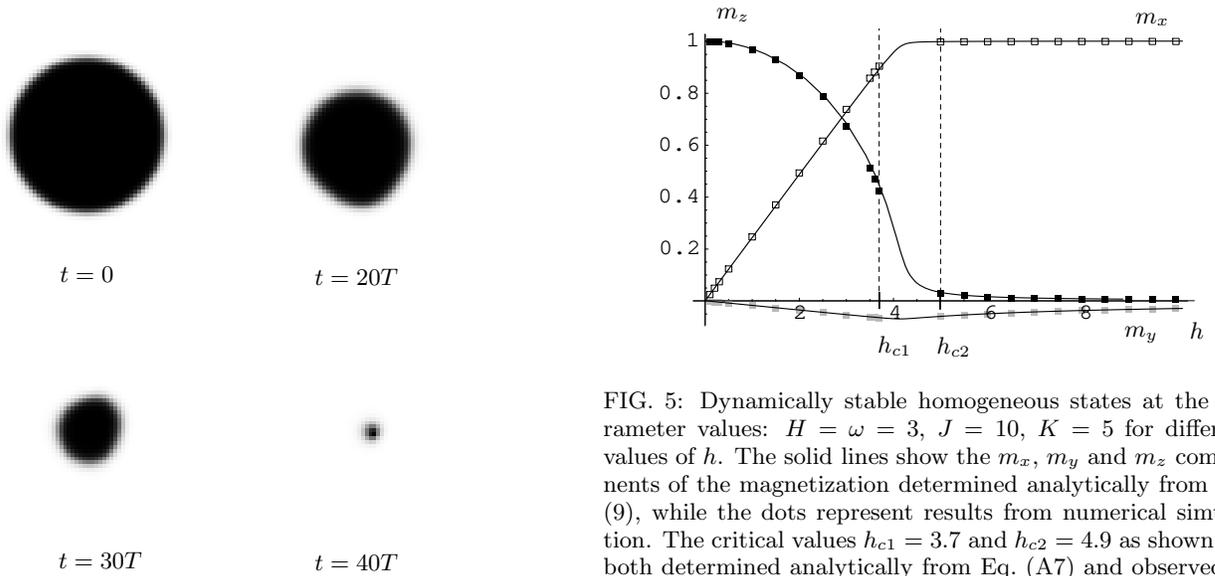}
\caption[]{Disintegration of a spin-up bubble in time
for a driving field amplitude just below the critical amplitude $h_{c3}$} \label{bubmovie}
\end{figure}

As indicated in Figure \ref{ndd3a}, there is tristability in the system, and which
pattern actually occurs depends on the initial conditions of the system.
Two hysteresis loops occurring in the system are shown in Figure \ref{ndd3}.
In particular, the behavior of the system at $h_{c3}$ is of interest. We select as
an initial state a domain state which would be stable just above
$h_{c3}$. Reducing $h$ a little below the critical value $h_{c3}$,
the two walls on either side of the ``spin-down'' domain begin to
move towards each other, and the magnetization within the walls to rotate faster and
faster than the angular frequency of the driving field. Eventually the walls
collide and leave the system in a homogeneous ``spin-up'' state. This process is shown in
Figure \ref{bubmovie}. The velocity of
walls in this transient state is smaller the closer we are to the value
$h=h_{c3}$, and it can take up to $60 T$, where $T=2\pi/\omega$ is
the period of the driving field, for the wall collision to occur.

\begin{figure}
\psfrag{x}{$m_x$}
\psfrag{y}{$m_y$}
\psfrag{z}{$m_z$}
\psfrag{h}{$h$}
\psfrag{a}{$h_{c1}$}
\psfrag{b}{$h_{c2}$}
\includegraphics[width=0.4\textwidth]{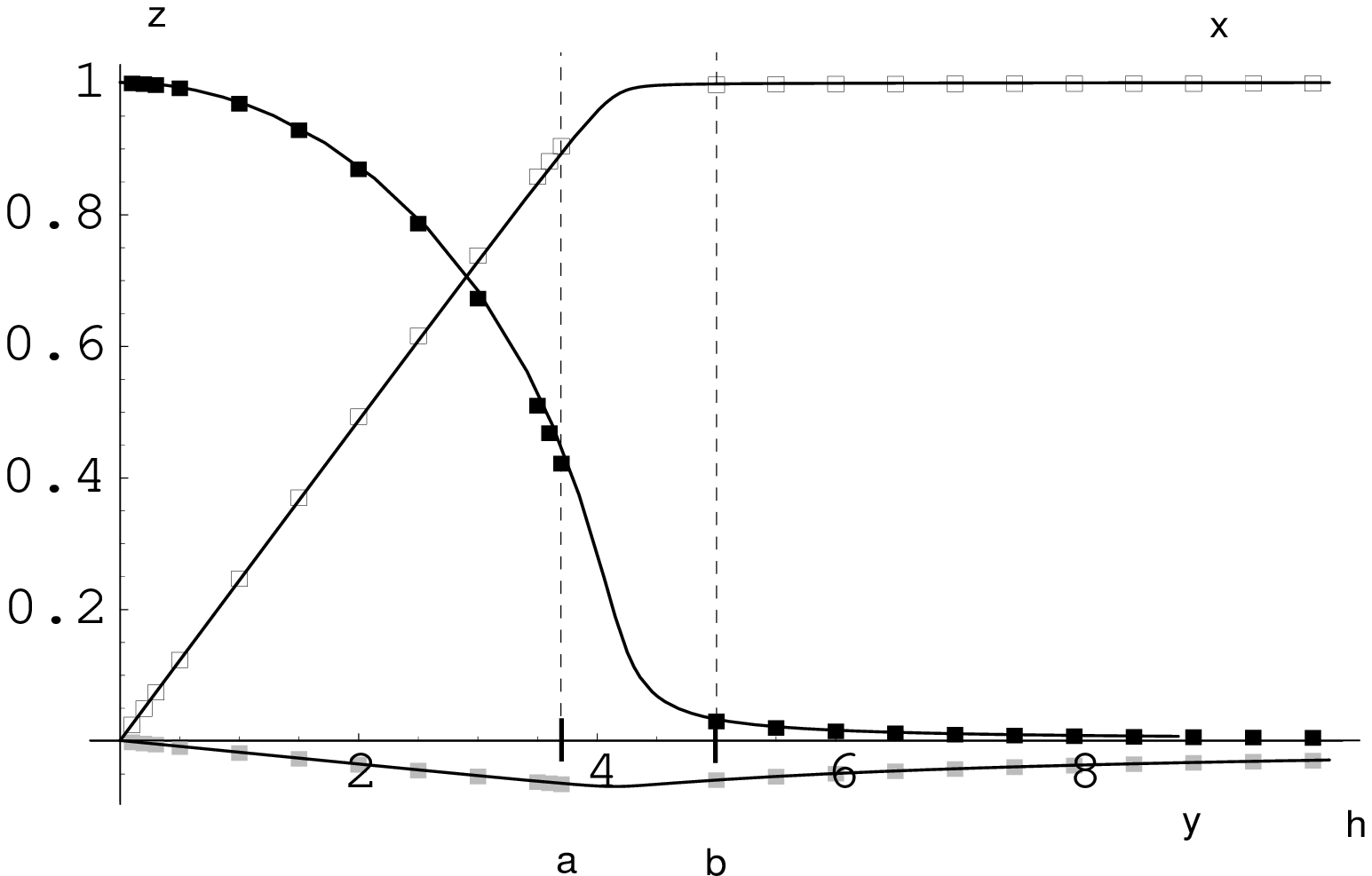}
\caption[]{Dynamically stable homogeneous states at the
parameter values: $H=\omega=3$, $J=10$, $K=5$ for different values of $h$. The
solid lines show the $m_x$, $m_y$ and $m_z$ components of the magnetization
determined analytically from Eq.\ (\ref{hfp}), while the dots represent results from
numerical simulation. The critical values $h_{c1}=3.7$ and $h_{c2}=4.9$
as shown are both determined analytically from Eq.\ (\ref{homoeqcrit})
and observed in numerical simulations}
\label{abb9.2} \end{figure}

\section{Analytical Investigation}

\subsection{Homogeneous Solution}\label{homosec}

Homogeneous solutions must satisfy the following
conditions
\begin{equation*}
\nabla^2\vec{m}=0 \quad ,\overline{m}_z=m_z \quad .
\end{equation*}
and so the effective magnetic field becomes:
\begin{equation}\label{hheff}
\vec{h}_{\rm eff}=H\vec{e}_z + h\vec{e}_x  + (K-1) m_z \vec{e}_z
\quad .
\end{equation}
We insert (\ref{hheff}) into equation (\ref{ll5}) and note that
for steady state solutions ($\partial_t \vec{m}=0$)
the expression in square brackets in (\ref{ll5}) must be parallel to
$\vec{m}$, with some unknown proportionality factor $\mu$:
\begin{equation}\label{h3}
h\vec{e}_x+(\delta+(K-1) m_z)\vec{e}_z+\gamma \vec{m}\times
\vec{e}_z=\mu \vec{m} \quad ,
\end{equation}
where we have introduced the following definitions:
\begin{equation}\label{h4}
\delta=H-\frac{\omega}{1+\Gamma^2}\quad ,\quad
\gamma=\frac{\Gamma \omega}{1+\Gamma^2} \quad .
\end{equation}
The parameter $\delta$ is a measure of the deviation of the system
from resonance and is called the detuning, while $\gamma$ is the
rescaled frequency.

Taking the scalar product of equation (\ref{h3}) with $\ex$, $\ey$ and $\ez$
respectively and eliminating the unknown factor $\mu$ yields two equations
for the 3 components of $\vec{m}$. A third equation is obtained from the
requirement that $|\vec{m}|=1$. We obtain a fourth order polynomial in $m_z$:
\begin{eqnarray}\label{hfp}
&&\left[(K-1)^2+\gamma^2\right] m_z^4+2 \delta (K-1)
m_z^3+\left[h^2+\delta^2\right. \nonumber\\
&& \left.-(K-1)^2-\gamma^2\right]m_z^2
-2 \delta (K-1) m_z-\delta^2=0\quad .
\end{eqnarray}
Either 2 or 4 solutions to equation (\ref{hfp}) are found,
depending on the parameter regime. A saddle-node bifurcation separates the
region of two solutions from that with four solutions.

The stability of these four fixpoint solutions is discussed by
means of a linear stability analysis (see Appendix).
Figure \ref{abb9.2} shows the components of the magnetization for that fixpoint
where $m_z > 0$, and the positions of $h_{c1}$ and $h_{c2}$, between which
this fixpoint is unstable to nonuniform perturbations with wavenumber $k \to 0$. For comparison, the
dynamically stable homogeneous states found numerically are also shown.

\subsection{Domain Solution}\label{domsec}

We make a simple ansatz to describe the magnetization in a two-domain state:
\begin{equation}\label{dan}
\begin{split}
\vec{m}(\xi)=\left\{\begin{array}{l l}
            \vec{m}^{(1)} &   0 < \xi < L q\\
            \vec{m}^{(2)} &  L q < \xi < L (1-q)
            \end{array}\right.
\end{split}
\end{equation}
i.e. we consider only the magnetization within each domain $i=1,2$, and imagine
ourselves to be  a
considerable distance away from the domain walls. The quantity $q$ describes the
proportion of the system in each domain.

\begin{figure}[t]
\psfrag{m}{$m_z$}
\psfrag{h}{$h$}
\psfrag{a}{$h_{c3}$}
\psfrag{b}{$h_{c4}$}
\psfrag{1}{\scriptsize 1}
\psfrag{2}{\scriptsize 2}
\psfrag{3}{\scriptsize 3}
\psfrag{4}{\scriptsize 4}
\psfrag{5}{\scriptsize 5}
\includegraphics[width=0.4\textwidth]{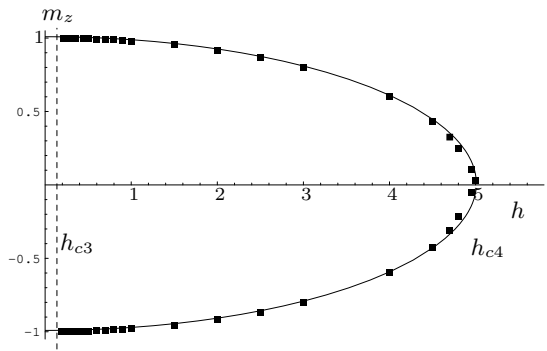}
\caption[]{The $z$-component of the magnetization within
dynamic domains as obtained from numerical simulations and analytical
calculation for different values of $h$.
Each pair of dots for a given value of $h$ corresponds to the $m_z$
values far from the domain wall for a single numerically stable
simulation; the solid lines show the analytical result Eq.\ (\ref{eqnd.4}). The upper bound
$h_{c4}=5.0037$ is both calculated from Eq.\ (\ref{upbound}) and observed numerically, while the
lower bound $h_{c3}=0.2$ is only observed numerically}
\label{abb9.3} \end{figure}

Again we look for steady state solutions ($\partial_t \vec{m}^{(i)}=0$), and demand
that the spatial inhomogeneity in each domain vanish ($\nabla^2 \vec{m}^{(i)}=0$).
The demand that the term in square brackets in (\ref{ll5}) be parallel to $\vec{m}$
yields:
\begin{equation}
-\he\ui+\frac{\omega}{1+\Gamma^2}\ez-\frac{\Gamma
\omega}{1+\Gamma^2}\vec{m}\ui\times\ez=\mu\ui\vec{m}\ui
\label{dgll}
\end{equation}
with
\begin{equation}\label{heffi}
\vec{h}_{\rm eff}\ui=H\vec{e}_z + h\vec{e}_x - \overline{m}_z
\vec{e}_z + K m_z\ui \vec{e}_z \quad .
\end{equation}
These two equations for $\vec{m}^{(1)}$ and $\vec{m}^{(2)}$ are
not independent, but rather are coupled via the demagnetizing term
$\overline{m}_z$ which holds for the entire system and appears in
each effective field $\vec{h}_{\rm eff}\ui$. Using the ansatz
(\ref{dan}) we write down the demagnetizing term as
\begin{equation}\label{davmz}
\overline{m}_z=\frac{1}{L}\int\limits_0^{L} m_z(\xi) {\rm d} \xi=  q \mza+ (1-q) \mzb \quad .
\end{equation}
A further ansatz still needs to be made to obtain domain
solutions, and motivated by the symmetry of the numerical results we set:
\begin{equation}\label{dm}
\begin{array}{c c c c c}
m_x^{(1)}  = & m_x^{(2)} & =: & m_x\quad ,\\
m_y^{(1)}  = & m_y^{(2)} & =: & m_y \quad ,\\
\mza  = & -\mzb  && \quad .
\end{array}
\end{equation}
Although this symmetry is not
inherent in the equations of motion, it is observed in all numerical simulations without
exception and,
more importantly, it fixes $q$ and permits a simple closed solution form.

By taking the scalar product of (\ref{dgll}) for $i=1,2$ with $\ex$, $\ey$ and $\ez$
and eliminating the unknown $\mu\ui$, we find consistency only if
\begin{equation}\label{d79}
\overline{m}_z = \delta \quad .
\end{equation}
This is an equilibrium condition for dynamic domains that fixes the proportion
of the system in each domain. The effective magnetic field (\ref{heffi})
in each domain $i$ is now
\begin{equation}\label{dheff}
\vec{h}_{\rm eff}\ui=(H-\delta)\vec{e}_z + h\vec{e}_x + K m_z\ui
\vec{e}_z \quad .
\end{equation}
Comparing equation (\ref{dheff}) with (\ref{hheff}), we see that
all results from the homogeneous calculation in Section
\ref{homosec} can now be applied, providing $H$ is replaced by
$H-\delta$ (thus $\delta$ is replaced by
$(H-\delta)-\frac{\omega}{1+\Gamma^2}=0$) and $K$ by $K+1$. Equation (\ref{hfp})
then becomes
\begin{equation}\label{dfixp}
m_z^{(i)^4}(K^2+\gamma^2)+m_z^{(i)^2}(h^2-K^2-\gamma^2)=0
\end{equation}
with solutions
\begin{equation}\label{eqnd.4}
m_z\ui=\pm\sqrt{1-\frac{h^2}{K^2+\gamma^2}}\quad .
\end{equation}
The other components of the magnetization are also simply expressed as
\begin{equation}\label{xyzd}
m_x =\frac{h K}{K^2+\gamma^2} \quad , \quad
 m_y =-\frac{\gamma h}{K^2+\gamma^2} \quad .
\end{equation}
Figure \ref{abb9.3} shows the $z$-component of the magnetization within each domain, as
given in (\ref{eqnd.4}) and compared to numerical simulations. We note that the domain solution can
only exist for
\begin{equation}\label{upbound}
h < \sqrt{K^2+\gamma^2} \quad .
\end{equation}
A linear stability analysis
(see Appendix) shows that the simple domain state (\ref{dan})
is stable for {\it all} values of $h$ where
the solution exists. This
is in contrast to the numerical evidence, which indicates that there is a lower bound (shown
as $h=h_{c3}=0.2$ in Figure \ref{abb9.3}) below which no numerical domain solutions exist. There is
{\it no} indication of such a lower bound in any calculations carried out using
the ansatz (\ref{dan}).

We recall that the ansatz (\ref{dan}) does {\it not}
describe an entire dynamic domain, only that part far from the domain wall.
Therefore any criteria for stability determined in performing a linear stability
analysis on (\ref{dan}) will be only {\it necessary} conditions for stability,
and not {\it sufficient} conditions. Critically, we have neglected any description of the shape or
dynamics of the domain wall, and merely assumed a discontinuous transition from one domain
to the next. In order to obtain a full description of dynamic domains, and, in particular,
to determine the position of the lower bound $h_{c3}$, it will be necessary to include the wall and
consider its effect on the stability of the system.

\section{Conclusions}

This work presents a model for a ferromagnetic film in which the
exchange, dipolar and anisotropy effects are taken into account.
Numerical simulation shows, depending on the amplitude of the
transverse pump field applied, different coexisting homogeneous
states and dynamic domain structures. The homogeneous state and
its stability is wholly understood, and, using a simple model,
certain features of the dynamic domain state are also clarified.
The magnetization within the domains is determined, as is the
proportion of the system in each dynamic domain. Nevertheless a
lower bound for the existence of dynamic domains is observed in
the numerical simulations that is not found with the simple
domain-state model.

An extension to this work has to include the effect of the domain
wall, both its structure and its dynamics. Such an extension has
been worked out \cite{Mayes} and will be published separately
\cite{mps} leading to an  understanding of the numerical results
in their entirety.
\begin{acknowledgments}
We wish to acknowledge interesting discussions with H. D\"otsch
and T. W\"obbeking
\end{acknowledgments}
\appendix

\section{Linear Stability Analyses}

\subsection{Homogeneous Solution}\label{hstabsect}

We denote
the basic solution computed above as $\vec{m}_0$. We then add a small time and space dependent
perturbation to $\vec{m}_0$, in particular we consider
one Fourier mode of such a perturbation:
\begin{equation}\label{han}
\vec{m}(\xi, t)=\vec{m}_0 + \delta \vec{m}(\xi,t)=\vec{m}_0+\varepsilon \vec{m}_1(t) e^{i k \xi}
\quad .
\end{equation}
Now the magnetization $\vec{m}(\xi,t)$ must have magnitude 1. Thus $\vec{m}_1$
must lie in a plane perpendicular to $\vec{m}_0$:
\begin{equation}\label{perpcrit}
\vec{m}_0 \cdot \vec{m}_1 =0 \quad .
\end{equation}
We insert ansatz (\ref{han}) into the Landau-Lifshitz equation in the rotating
reference frame (\ref{eq1}) and sort for powers of $\varepsilon$. To order $\varepsilon$:
\begin{eqnarray}\label{ho1}
-\dot{\vec{m}_1}&=&\vec{m}_1\times (\vec{h}_{\rm eff,0}-\omega
\vec{e}_z)+ \vec{m}_0\times\vec{h}_{\rm eff,1}+\nonumber\\
&&\Gamma\left[ \vec{m}_1\times (\vec{m}_0\times \vec{h}_{\rm
eff,0})+ \vec{m}_0\times (\vec{m}_1\times \vec{h}_{\rm
eff,0})\right.+\nonumber\\
&&\quad\left.\vec{m}_0\times (\vec{m}_0\times \vec{h}_{\rm eff,1})
\right] \quad ,
\end{eqnarray}
where we have introduced
\begin{eqnarray*}
\vec{h}_{\rm eff,0}&=&H\vec{e}_z + h\vec{e}_x + (K-1) m_{z,0} \vec{e}_z\quad ,\\
\vec{h}_{\rm eff,1}&=&-J k^2 \vec{m}_1 e^{i k \xi} +K m_{z,1}
\vec{e}_z e^{i k \xi} -\overline{\delta m_z}\,\vec{e}_z \quad .
\end{eqnarray*}
The field $\vec{h}_{\rm eff,1}$ depends on whether the case $k=0$
or $k\neq 0$ is being considered, i.e. whether the perturbation $\delta\vec{m}$ is
uniform or non-uniform. If $k =0$, $\overline{\delta m_z}=m_{z,1}$,
whereas if $k \neq 0$, the fluctuations averaged over the entire
system are zero and so $\overline{\delta m_z}=0$. Therefore
\begin{eqnarray*}
\mbox{for} \quad k=0\,, &\qquad & \vec{h}_{\rm eff,1}= (K-1) m_{z,1} \vec{e}_z\quad ,\\
\mbox{for} \quad k\neq 0\,, &\qquad & \vec{h}_{\rm eff,1}= K
m_{z,1} \vec{e}_z e^{i k \xi}-J k^2 \vec{m}_1 e^{i k \xi}\quad .
\end{eqnarray*}
We introduce a coordinate system defined as
\begin{equation}\label{hcoord}
\vec{e}_1=\vec{m}_0,\,
\vec{e}_2=\vec{m}_0\times \vec{e}_z, \,
\vec{e}_3=\vec{m}_0\times(\vec{m}_0\times \vec{e}_z) \quad .
\end{equation}
In this coordinate system $\vec{m}_1$ is written as
\begin{equation*}
\vec{m}_1(t)=\beta(t) \vec{e}_2+\gamma(t) \vec{e}_3
\end{equation*}
and equation (\ref{ho1}) assumes the form
\begin{eqnarray}\label{eq3}
-\dot \beta \vec{e}_2 - \dot \gamma \vec{e}_3 &= &
(\vec{h}_{\rm eff,0}\cdot \vec{m}_0 -\omega m_{z,0})(-\beta
\vec{e}_3 +\gamma \vec{e}_2)+\nonumber\\
&& \Gamma(\vec{m}_0 \cdot \vec{h}_{\rm
eff,0}) (\beta \vec{e}_2+\gamma \vec{e}_3)+\nonumber\\
& &\vec{e}_1\times \left[\vec{h}_{\rm
eff,1}+\Gamma \vec{e}_1\times  \vec{h}_{\rm eff,1} \right] \quad .
\end{eqnarray}
We examine the four fixpoint solutions of (\ref{hfp}) for stability
with respect to uniform ($k=0$) and nonuniform ($k\neq 0$) perturbations by considering
the eigenvalues of the linear operator  $\underline{\underline L}$
defined by the right-hand side of equation (\ref{eq3}).

The eigenvalues of the $2 \times 2$ matrix $\underline{\underline L}$ for $k=0$ are positive
for two of the four fixpoints. For the other two fixpoints,
the trace of $\underline{\underline L}$ for $k \neq 0$
is negative for all values of $k$, so we only need consider its determinant, given by
\begin{equation}\label{hdt}
Det(L_{k\neq 0})=(1+\Gamma^2)\left(k^4 J^2 + k^2 J P + Q\right)\quad ,
\end{equation}
where
\begin{eqnarray*}
P&=& \left(K (m_{z,0}^2-1)+2 \vec{m}_0\cdot \vec{h}_{\rm
eff,0}\right)-2 \frac{\gamma}{\Gamma} m_{z,0} \quad ,\\
Q&=& (\vec{m}_0\cdot \vec{h}_{\rm
eff,0}) \left(\vec{m}_0\cdot \vec{h}_{\rm
eff,0} + K (m_{z,0}^2-1) \right)\\
&&+ \frac{\gamma}{\Gamma} m_{z,0} \left(\vec{m}_0\cdot \vec{h}_{\rm
eff,0}+ \omega m_{z,0}\right) \quad .
\end{eqnarray*}
Varying $h$, the determinant first becomes negative (and hence the fixpoint unstable) for $k\to 0$
at critical values of the driving field called $h_{c1}$ and $h_{c2}$ given by
\begin{equation}\label{homoeqcrit}
P-\sqrt{P^2-4 Q}=0 \quad .
\end{equation}

\subsection{Domain Solution}

The stability analysis of the domain solution is performed analogously to that
of the homogeneous solution in Section \ref{hstabsect}. A small perturbation
is added to the magnetization in each domain $i=1,2$. As the Fourier modes in the first domain
would couple to all of those in the second domain, we instead perform the calculation in real space,
writing
\begin{equation*}
\vmui(\xi,t)=\beta(\xi,t)\ui\vec{e}_2^{(i)}+\gamma(\xi,t)\ui\vec{e}_3^{(i)}\quad .
\end{equation*}
We obtain a system of 2 equations for each $(\beta\ui,\gamma\ui)$, equivalent
to (\ref{eq3})
\begin{eqnarray}\label{dll2}
&&-\dot \beta\ui \vec{e}_2\ui - \dot \gamma\ui \vec{e}_3\ui  = \nonumber\\
&&\quad \left[\vmo\ui \cdot \vec{h}_{\rm eff,0}\ui - \omega m_{z,0}^{(i)}
\right](-\beta\ui \vec{e}_3\ui +\gamma\ui \vec{e}_2\ui)\nonumber \\
&&\quad+
\Gamma(\vmo\ui \cdot \vec{h}_{\rm eff,0}\ui) (\beta\ui
\vec{e}_2\ui+\gamma\ui \vec{e}_3\ui)\\
&&\quad +\left[ K (m_{z,0}^2-1)\gamma\ui -
(\overline{m}_z)_1 \right]
(\vec{e}_2\ui+ \Gamma \vec{e}_3\ui) \nonumber\\
&&\quad +J \left[\beta{''}\ui (\vec{e}_3\ui-\Gamma
\vec{e}_2\ui)+\gamma{''}\ui (-\vec{e}_2\ui - \Gamma \vec{e}_3\ui)\nonumber
\right]
\end{eqnarray}
where $\vec{h}_{\rm eff,0}^{(i)}$ is defined in (\ref{dheff}).

In the case of {\it uniform} perturbations, equation (\ref{dll2}) reduces to a system of
four coupled ordinary differential equations. Numerically computing the eigenvalues of this
$4\times 4$ matrix shows that all eigenvalues are negative for
$h < h_{c4} = \sqrt{K^2+\gamma^2}$
and so the domain state
is stable to (local) uniform perturbations.

For {\it non-uniform} perturbations, we descritize equation
(\ref{dll2}) by replacing the
spatial derivatives with finite difference expressions. For all values of
$h < h_{c4} = \sqrt{K^2+\gamma^2}$, the real part of the eigenvalue spectrum is strictly negative
 for both solutions $m_z^{(i)}$ in (\ref{eqnd.4}).
Hence the domain state is also  stable to (local) non-uniform perturbations.

Finally, we may consider what happens when the {\it position of the wall is displaced from its
equilibrium position} given by (\ref{d79}), i.e., additionally letting
$q \to q_0 + \varepsilon q_1(t)$. Again, we obtain a linearized equation of motion. Integrating
this equation over a small section of the containing the wall, and then letting the
size of this section go to zero
\begin{equation*}
\lim_{\overline\varepsilon\to 0} \int\limits^{L
q_0+\overline\varepsilon}_{L q_0-\overline\varepsilon} \ldots
\,\,{\rm d}\xi \quad ,
\end{equation*}
we obtain a single equation for $q_1(t)$:
\begin{equation}\label{qeqn}
\dot q_1 = -K \left[1-\Gamma(1-m_{z,0})\right] q_1 \quad .
\end{equation}
The expression in square brackets in (\ref{qeqn}) is strictly
positive, and so again the domain solution is stable to
displacement of the wall from its equilibrium position.


\end{document}